\documentclass[journal=apchd5,manuscript=article]{achemso}

\usepackage[version=3]{mhchem} % Formula subscripts using \ce{}
\usepackage{xr}
\usepackage{comment}
\usepackage{upgreek}
\usepackage{lineno}
\usepackage{enumitem}

\usepackage{xcolor}

\newcommand\aZPL{75.2\%~}
\newcommand\aZPLAR{83.0\%~}
\newcommand\aZPLARPEC{98.9\%~}  % for ZPL with PEC and AR, optimal distance is 550 nm.
\newcommand\aPSBARPEC{77.3\%~}  % for PSB with PEC and AR, optimal distance is 300 nm.
\newcommand\bNV{65.9\%~}   % scheme b for whole NV spectrum 600-800nm, weighted by NV spectrum, in diamond baseline is 73.45%
\newcommand\bNVAR{69.4\%~}
\newcommand\bNVARPEC{82.2\%~}

\newcommand\aBaseline{0.36\%~}
\newcommand\bBaseline{12.63\%~}

\author{Jiabao Zheng}
\email{jz2466@columbia.edu}
\affiliation{Department of Electrical Engineering, Columbia University, New York, New York 10027, USA.}
\alsoaffiliation{Department of Electrical Engineering and Computer Science, Massachusetts Institute of Technology, Cambridge, Massachusetts 02139, USA.}

\author{Andreas C. Liapis}
\affiliation{Center for Functional Nanomaterials, Brookhaven National Laboratory, Upton, New York 11973, USA.}

\author{Edward H. Chen}
\affiliation{Department of Electrical Engineering and Computer Science, Massachusetts Institute of Technology, Cambridge, Massachusetts 02139, USA.}

\author{Charles T. Black}
\affiliation{Center for Functional Nanomaterials, Brookhaven National Laboratory, Upton, New York 11973, USA.}

\author{Dirk Englund}
\email{englund@mit.edu}
\affiliation{Department of Electrical Engineering and Computer Science, Massachusetts Institute of Technology, Cambridge, Massachusetts 02139, USA.}

\title{Near-unity collection efficiency from quantum emitters in bulk diamond using chirped circular dielectric gratings}
\abbreviations{IR,NMR,UV}

\begin{document}


\begin{abstract}
Efficient collection of fluorescence from nitrogen vacancy (NV) centers in diamond underlies the spin-dependent optical read-out that is necessary for quantum information processing and enhanced sensing applications. 
The optical collection efficiency from NVs within diamond substrates is limited primarily due to the high refractive index of diamond and the non-directional dipole emission.
Here we introduce a light collection strategy based on chirped, circular dielectric gratings that can be fabricated on a bulk diamond substrate to redirect an emitter's far-field radiation pattern. Using a genetic optimization algorithm, these grating designs achieve \aZPLARPEC collection efficiency for the NV zero-phonon emission line, collected from the back surface of the diamond with an objective of aperture 0.9.
Across the broadband emission spectrum of the NV (600-800 nm), the chirped grating achieves \bNVARPEC collection efficiency into a numerical aperture of 1.42, corresponding to an oil immersion objective again on the back side of the diamond. Our proposed bulk-dielectric grating structures are applicable to other optically active solid state quantum emitters in high index host materials.
\end{abstract}


\section*{Introduction}

The negatively charged nitrogen vacancy (NV) center in diamond has emerged as a leading semiconductor quantum system with well-controlled electronic orbitals and spin states. A number of applications are being developed, including quantum computation \cite{SimonBenjamin2014PRX, NemotoPRX2014}, quantum networking~\cite{15NHansonBelltest}, quantum simulation\cite{14PRBJiangfengNV-TI}, quantum error correction~\cite{16NcommTaminuauQECNV} and precision sensing\cite{12NanoLettWrachtrupNV-Tsensing, Degen2016SensingReview, 15NcommCarlosNVscanningThermalConImage}.  There has also been considerable progress in developing other atom-like quantum emitters in solids with unique features.\cite{17NcommPingaultSiVCoherentControl, 11PRBBecherSiVlinewidth,14AtatureNcommSiVspinphoton, 14HeppPRLSiVElevel, 14PingaultPRLSiVCPT,
04NJPWrachtrupNE8, 15NsrIwasakiGeV, 10PRAAharonovichCrRelatedV, 15NmatWidmannSiC} 

The efficient collection of fluorescence emitted by individual quantum emitters is of central importance in advancing these quantum technologies. For instance, the photon collection efficiency $\eta$ of the NV's total emission (zero phonon line and phonon side-band) determines the minimum detectable magnetic field with a single NV as $|\delta \mathbf{B_{min}}| \propto 1/ \sqrt{\eta}$~\cite{08NPhysTaylorNVsensing, 07NPhysBudkerNVmagReview, 12MRSBullHongNVmagReview}, assuming photon shot-noise limited detection. The total NV fluorescence collection efficiency is also essential for fast syndrome measurements in quantum error correction~\cite{16NcommTaminuauQECNV}. In quantum networking, the rate at which two NV centers can be optically entangled depends on how efficiently two identical photons can be detected, so that the rate is proportional to $ \eta ^2$. Higher collection efficiency is necessary to allow the entanglement generation rate to exceed the decoherence rate of the NV.   

Efficient collection of NV fluorescence remains challenging and is hindered primarily by two factors: (1) The dipolar emission pattern of NV fluorescence\cite{05NPhyAwschalomNVdipole} is challenging to efficiently collect with a single port \cite{12PRBNV4PortCollection}; (2) The large refractive index mismatch between diamond ($n =$ 2.41 at $\lambda = $ 637 nm) and air confines the emitted photons by total internal reflection (TIR), allowing only a small fraction of the emission to couple out. These two reasons set the collection efficiency of a point source (shown in Fig.\ref{BullseyeIllustration}(a)) to a baseline value of $\frac{1}{2}\int^{\theta_{\text{TIR}}}_{0} \text{sin} \theta ~ \text{d}\theta \times 0.8 \sim 3.6\%$, where 0.8 is the approximate averaged Fresnel transmission coefficient for diamond-air interface within the TIR window. (See Supporting Information section S2 for a rigorous calculation on the baseline values of collection efficiency for a dipole emitter).  In addition, the NV photon generation rate is relatively low mainly because of its rather long excited state lifetime ($\sim$ 13 ns) \cite{13DohertyPRNVreview}; (See Supporting Information section S1 for estimation of NV$^-$ intrinsic emission rate). This lifetime could be reduced using the Purcell effect.  However, simply enhancing the NV emission rate via Purcell effect is likely to decrease the spin readout contrast, as recently discussed theoretically\cite{15PRBNVPurcellSNR} and experimentally \cite{17BogdanovNVcontrastPurcell}.

To tackle those challenges, various optical structures have been investigated, including tapered fibers\cite{16LSA_Rishi_NVcollectionTaperFiber}, metallic mirrors placed in close proximity to NV \cite{14JOAndersenNVSilverMirror}, nanofabricated solid immersion lenses \cite{SILBrien, SILHadden}, parabolic mirrors\cite{14BensonAPLParabolicMirrorND}, nanowire arrays \cite{10NnanoLoncarNanoWire},  dielectric optical antennas\cite{opticalNVantenna}, hybrid waveguide structures \cite{15NLWrachtrupDiamondPost}, diamond-silver apertures\cite{choy2011enhanced} and plasmonic gratings \cite{SiO2Grating}.

In this work, we consider bullseye gratings that consist of a series of concentric circular rings dry-etched into diamond with an NV in the center. Circularly symmetric structures are generally preferred because they eliminate the need for alignment to the dipole orientation of the emitter. Circular gratings have previously been considered for cavity-defect mode enhancement\cite{scheuer2006annular}, high-Q resonators \cite{04.PRE.Scheuer.Bullseye}, and emission out-coupling of quantum dots\cite{11APLSrinivasanBullseyeQD, IEEE.Srinivasan.BullseyeQD}. Experiments with circular gratings on a diamond membrane have shown measured single NV photon count rates of $\sim$ 3 Mcps  \cite{15NanoLett.Li.MemBullseyeNV}.

An important challenge in imposing a grating on a bulk dielectric surface relates to the pitch of the air slits. In a thin membrane supporting only the fundamental mode in the vertical direction, a constant-period pitch works well\cite{11APLSrinivasanBullseyeQD, 15NanoLett.Li.MemBullseyeNV}. For a bulk dielectric medium, however, the grating needs to accommodate a range of $k$-values in the out-of-plane direction. In this case, a non-periodic, chirped grating provides better coupling into a target out-of-plane propagating mode. The chirped grating approach can also improve the coupling efficiency over a broad spectral range, whereas constant-period gratings are optimal for narrowband operation. Chirped optical structures have been developed for applications that require broad bandwidth and wide angular range operations, such as broadband mirrors for ultrafast lasers\cite{1997Kartner}, or ultra-thin photovoltaic devices with enhanced absorption\cite{Agrawal2008q,14AcsPhotonicsZhengChirpedDBR}.  Additionally, implementing the chirped circular structure on a relatively thick diamond film ($\sim$ 20 $\mu$m) instead of on a single mode diamond membrane ($\sim$ 200 nm) could further improve the collection efficiency by index guiding\cite{11NphotonicsGotzingerNearUnity}. Finally, a mirror can be placed on the top side of the grating to reflect the upwards-propagating emission and help cover a complete range of $k$-values.

The chirped circular grating structures introduced here enable near-unity collection efficiency into the detection apparatus for different numerical aperture (NA). These bulk-dielectric grating structures can be fabricated in bulk diamond and do not require diamond membranes, which are not commercially available and difficult to fabricate\cite{14HodgesNJPDiamondMembrane, 16PrawerNanoLettDiamondMembrane, 16BecherPSSDiamondMembrane}. We numerically evaluate the performance of chirped circular gratings optimized for three specific detection schemes:

\begin{enumerate}[label=(\alph*)]
\item Collection of the NV's zero-phonon line (ZPL) through an objective lens under cryogenic conditions, as required for photonic entanglement of remote NV centers;
\item Collection of the phonon side band (PSB) through an objective lens under cryogenic conditions, as required for quantum state estimation and quantum error correction; 
\item Collection of the NV's ZPL and PSB through an oil immersion objective lens at room temperature, as required for many quantum-enhanced sensing applications. 
\end{enumerate}

For detection scheme (a), the chirped bulk-dielectric grating achieves a near-unity collection efficiency of \aZPLARPEC into an NA of 0.9 air objective (Bullseye grating A), corresponding to a $\sim$ 275 fold improvement in collection efficiency compared to unpatterned diamond. For scheme (b), we predict a collection efficiency of 68.8\% for the broad PSB emission with same Bullseye grating A . In scheme (c), a chirped circular grating optimized for broadband efficient collection (Bullseye grating B) yields an averaged collection efficiency of \bNVARPEC into NA = 1.42 and averaged Purcell factor of 1.18.  We also examine the sensitivities of the collection efficiency enhancements for NV displacement from the optimal position. For grating A, the enhancement stays within 15.9\% of the maximum as long as the NV is within 20 nm from its ideal position. Our chirped circular grating promises high collection efficiency with moderate Purcell factors for emitters embedded in a high index host material. The chirped circular grating is flexible in design for different experimental conditions, has negligible photon loss compared to plasmonic gratings, is relatively robust to NV position errors, and can be applied to other atomic semiconductor quantum emitters\cite{17NcommTimSiVFIB}, such as other solid state defects or quantum dots\cite{11APLSrinivasanBullseyeQD, IEEE.Srinivasan.BullseyeQD}.

\section*{Theoretical Considerations}

The chirped circular grating (`chirped bullseye structure')  consists of a series of concentric air slits in diamond surrounding the NV defect, as shown in Fig.\ref{BullseyeIllustration}(b) (perspective view), where the red dot indicates the position of the NV center.  An objective lens on the bottom of the diamond film collects the NV fluorescence. The width and position of each air slit are optimized to maximize the fluorescence emitted into the collection window (e.g. the numerical aperture of the objective lens). The collection efficiency is calculated through a far-field projection, taking into account the non-unity Fresnel transmission coefficient of the bottom diamond-air interface, and integrating the far-field power over the collection window. Experimentally, the far-field emission pattern can be imaged with a Bertrand lens at the back focal plane. \cite{novotny2012principles} We optimized the chirped bullseye structure using particle swarm optimization algorithm and the simulation is based on Finite-Difference Time-Domain (FDTD) method (Lumerical FDTD solutions). 

We discuss the appropriate figure of merit (FOM) for the problem of efficient collection of NV fluorescence. Because the photonic structure modifies the emitter's spontaneous emission rate via Purcell effect, the total collected photon number is proportional to the product of the Purcell factor $F_p$ and the photon collection efficiency $\eta$. 
However, optical readout of NV centers relies on their spin contrast in fluorescence intensity, which may be reduced for strong Purcell enhancement. S. A. Wolf \textit{et al.} discussed that the maximal SNR of the NV optical spin readout can be obtained for a rather low Purcell factor (optimal $F_p$ $\sim$  1-5 depending on excitation conditions), provided that the spin mixing process is radiative\cite{15PRBNVPurcellSNR}, which agrees with recent experimental investigation on the NV spin contrast under Purcell enhancement using plasmonic structures \cite{17BogdanovNVcontrastPurcell}. This finding indicates that one should aim at modifying the NV radiation pattern instead of the emission rate for higher SNR in the experiments. Therefore, we consider photon collection efficiency instead of absolute photon count rate as the FOM for optical spin readout in the three experimental schemes.

With collection efficiency denoted $\eta(\text{NA$_0$}, \lambda)$ set as the FOM for the discussion,  we now analyze the collection efficiency for NV in diamond.  $\eta = \text{P}_{coll}^{\downarrow}/\text{P}_{tot}$ and $\text{P}_{tot} = \text{P}_{rad} + \text{P}_{\text{NR}}  = \text{P}_{coll}^{\downarrow} +  \text{P}_{lost}^{\downarrow}  + \text{P}^{\uparrow} + \text{P}_{\text{NR}}$, where $\text{P}_{coll}^{\downarrow} = \int_{\text{NA}_{0}} \text{P}^{\text{far-field}}_{rad}(\theta,\phi)~\text{sin} \theta ~\text{d}\theta ~\text{d}\phi$ indicating the downwards power collected by an objective lens with $\text{NA} = \text{NA}_{0}$, $ \text{P}_{lost}^{\downarrow} $ is the photon emission rate outside of the collection $\text{NA}_{0}$, $\text{P}^{\uparrow}$ indicates the fluorescence from NV in the bullseye that emits upwards, and $\text{P}_{\text{NR}}$ is the emission power lost non-radiatively. $\text{F}_{p} = \text{P}_{rad}/\text{P}_{0}$  is the Purcell factor, where $\text{P}_{rad}$ is the radiative power from the NV center in the presence of the bullseye grating, and $\text{P}_{0}$ is the baseline radiative power from the NV center in a homogeneous diamond environment (related to NV intrinsic emission rate, see Supporting Information S1).

We will consider here three different application scenarios that require efficient light-collection under various experimental conditions: 

\begin{enumerate}[label=(\alph*)]
\item Heralded quantum entanglement between two remote NVs, where the zero phonon emission ($\lambda_{\text{ZPL}}$ = 637 nm) from each individual NV needs to be collected under cryogenic conditions with a high-NA dry objective lens (NA=0.9). Such NVs need to be deeply embedded inside the diamond for the NV to be sufficiently shielded from surface-related impurities and nearly transform-limited emission to be possible\cite{14NanoLettYiwen100nmImplantNV}. We consider 100 nm depth for a matching reference by Y. Chu \textit{et al.}\cite{14NanoLettYiwen100nmImplantNV}. 

\item Under the same experimental conditions as in (a), we also evaluate the performance of the bullseye grating in collecting the NV PSB for quantum error correction experiments.

\item For sensing external magnetic or electric fields, where the NV should be close to the diamond surface ($\leq$ 10 nm)\cite{12PRBDegenShallowNVSpin},  we consider applications in experiments performed at room temperature with an oil immersion lens. For this discussion, we assume an oil refractive index of 1.52 and an objective NA of 1.42 corresponding to the representative Olympus PLAN APO 60X OIL OB objective, and we choose 600-800 nm as the corresponding collection bandwidth of interest.  
\end{enumerate}

With the above discussion, the parameters in the simulation and resultant performance are summarized in table \ref{schemes}.

\begin{table}
\centering
\begin{tabular}{l*{7}{c}r}
scheme & (a) & (b) & (c) \\
\hline
dipole orientation & vertical & vertical & horizontal \\
dipole depth (nm)  & 100 & 100 & 10 \\
$\lambda$ (nm) & 637 & 640-800 & 600-800 \\
NA$_0$  & 0.9 & 0.9 & 1.42 \\
collection medium & air (n=1) & air (n=1) & oil (n=1.518) \\
design & Grating A & Grating A & Grating B \\
$\eta(\text{NA$_0$},\lambda)$ & \aZPLARPEC  &  \aPSBARPEC  &  \bNVARPEC  \\
$\eta_{\text{baseline}}$ & \aBaseline & \aBaseline & \bBaseline\\
\end{tabular}
\caption{Three experimental schemes for the Bullseye gratings}
\label{schemes}
\end{table}

\begin{comment}
\begin{tabular}{l*{7}{c}r}
scheme     & dipole & dipole depth ($nm$)  & $\lambda$ ($nm$)  & NA$_0$  & collection medium & design & $\eta(\text{NA$_0$},\lambda)$ & $\eta_{\text{baseline}}$\\
\hline
 (a) & vertical & 100 & 637 & 0.9   & air (n=1) & Grating A & 98.8\% & 0.3579\%\\ 
\hline
 (b) & vertical &100 & 640-800 & 0.9 & air (n=1) & Grating A & 68.8\% &  0.3579\%\\ 
\hline
 (c) & horizontal & 10 & 600-800 & 1.42 & oil (n=1.518) & Grating B & 82.2\% & \bBaseline\end{tabular}
\end{comment}

\section*{Results and Discussion}

\subsection*{Grating A for scheme (a) and (b)}

We first investigate for the collection efficiency of the NV ZPL into an NA of 0.9, which corresponds to a maximum collection angle of 21.9$^\circ$ in diamond. We optimize the circular grating structure to maximize $\eta$($\lambda$=637 nm, NA$_0$=0.9), assuming the NV electric dipole is perpendicular to the diamond-air interface. This is a nonideal dipole orientation for efficient fluorescence collection, since most of the emission is at high $k$ modes which are difficult to collect efficiently. We choose this dipole orientation to demonstrate the effectiveness of the bullseye grating. The optimized circular grating (Grating A) achieves collection efficiency $\eta$=\aZPL, corresponding to a $\sim$ 210 fold enhancement from the baseline value of $\eta$=\aBaseline. For comparison, the power emitted into the same collection window within diamond reaches 83.1\% of the total emission power. The difference in $\eta$ is due to the nonunity Fresnel transmission coefficient (average $\sim$ 80\%) of the bottom diamond-air surface. With an anti-reflection (AR) coating layer of index $n_{\text{AR}} = \sqrt{n_\text{diamond}n_{\text{air}}} ~ \sim 1.55$ and thickness = $\lambda_{\text{ZPL}}/4n_{\text{AR}} ~ \sim 102.6$ nm on the bottom diamond surface,  the averaged transmission for the bottom interface increases to $\sim$ 99\% while $\eta$ increases to \aZPLAR. 

\begin{figure}[htbp]
 \centering
  \includegraphics[width=17cm]{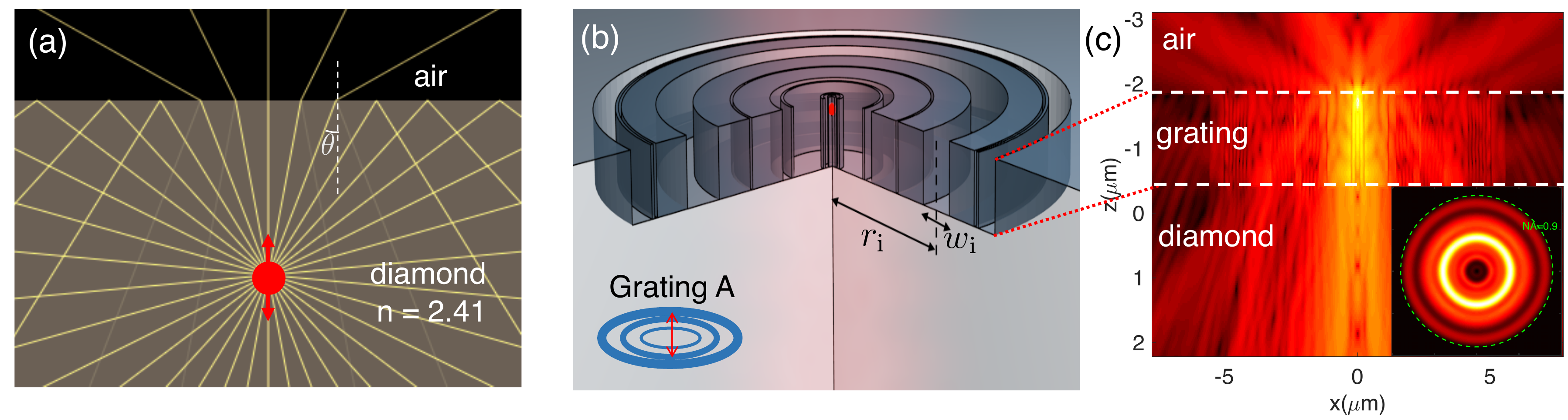}
\caption{(a) Ray optics simulation for diamond-air interface. Critical angle $\theta_{\text{TIR}}=\text{sin}^{-1}~ (1/n_{\text{diamond}}) $. (b) Chirped diamond grating, with NV location indicated as red dot. (c) Side view of optical mode $|\mathbf{E}|^2$ (in $log_{10}$ scale). Inset: far-field pattern of the ZPL emission from Grating A, the dashed green line shows the extent of NA = 0.9.} 
\label{BullseyeIllustration}
\end{figure}

\begin{figure}[htbp]
 \centering
  \includegraphics[width=16cm]{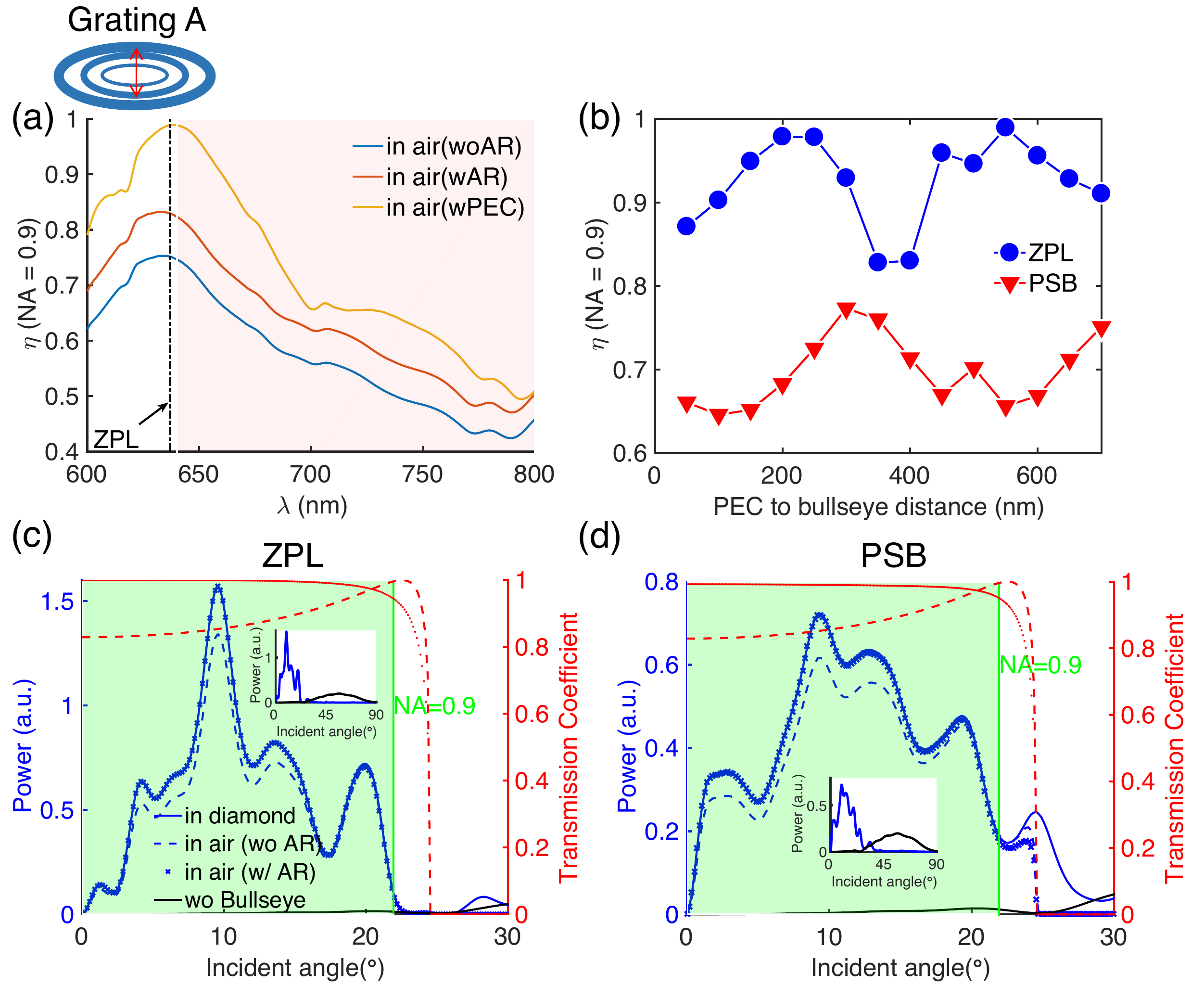}
\caption{\textbf{For Grating A:} (a) Broadband collection efficiency (NA = 0.9), without AR coating (blue line), with AR coating on bottom diamond surface (red line), and with 550 nm PEC-grating distance (yellow line). The pink shaded area indicates the wavelength range of PSB. (b) $\eta$($\text{NA$_0$}$=0.9) with PEC mirror over the top surface of Grating A at varying distance (with AR coating on the back diamond surface), for ZPL (blue) and PSB (red) emission. (c) Power angular dependance on the back diamond surface for ZPL emission. Left axis: Collected power in diamond medium (blue solid line), in air after transmission through bottom interface (blue dashed line), in air with an AR coating (blue crosses), in diamond medium but without Grating A (black solid line). The green shaded area indicates the collection window of an NA = 0.9 air objective. Inset: Collected power over 0-90$^{\circ}$ angular range. Right axis: Angular transmission coefficient for diamond-air interface without (red dashed line) and with (red solid line) an AR coating. (d) Same as (e), averaged power angular dependance over PSB.}
\label{BullseyeLT}
\end{figure}

The enhancement in collection efficiency is from the modification of far-field radiation profile to minimize $\text{P}_{lost}^{\downarrow}$. Fig.\ref{BullseyeIllustration}(c) inset shows the simulated far-field pattern projected to air for Grating A with AR coating. Note that both the dipole emitter and the grating, and the resulting far-field pattern, have the same circular symmetry. The mode in Bullseye A is \textit{p} polarized because the vertically polarized dipole source has only \textit{p} polarized emission. The angular power distribution of $p$ polarization component for Grating A with AR coating (Fig.\ref{BullseyeLT}(c)) indicates that bullseye structure effectively converts the nondirectional dipole emission (black solid line) into the collection window (plot in blue lines). Additionally, the AR coating improves the collection efficiency by maximizing the bottom interface transmission (plot in red lines) from an average value of $\sim$ 80\% to $\sim$ 99\%. 

Lastly, the collection efficiency can be further improved by placing a reflector over the grating to minimize $\text{P}^{\uparrow}$. The reflector could consist of an NSOM tip or silver coated fiber facet with piezo controlled stage \cite{14JOAndersenNVSilverMirror}. We find that when the reflector is 550 nm over the diamond surface, the collection reaches \aZPLARPEC. We assumed here a perfect reflector, modeled as perfect electric conductor in the simulation. The yellow line in Fig.\ref{BullseyeLT}(a) shows the spectrally resolved collection efficiency for this reflector position. 

We also evaluate the collection efficiency over the PSB with Grating A. We plot the power angular dependence for weighted averaged PSB emission in Fig.\ref{BullseyeLT}(d) and found similar enhancement with an AR coating layer. We also simulated with PEC reflector over the grating for PSB case, which is shown in Fig.\ref{BullseyeLT}(b), indicating that the optimal distance for PSB collection is 300 nm with $\eta$($\text{NA$_0$}$=0.9, $\lambda$=640-800 nm) = \aPSBARPEC.

 \subsection*{Grating B for scheme (c)}
For many quantum sensing applications, efficient collection over the whole NV spectrum is required. These experiments are performed at room temperature with high-NA oil immersion objectives. Here we set the FOM of the optimization to be the collection efficiency weighted by the NV emission spectrum and averaged over the range $\lambda$ = 600--800 nm. This we denote as
 $\bar{\eta}$($\text{NA$_0$}$=1.42, $\lambda$=600-800 nm) = $\eta($\text{NA$_0$}$,\lambda)\times\rho_{\text{NV}}(\lambda)$, where $\rho_{\text{NV}}(\lambda)$ is the normalized power spectral density of NV emission. We also assume the collection medium is n=1.518 oil and a horizontal dipole configuration, which corresponds to a NV in a (111) terminated diamond film.

\begin{figure}[htbp]
 \centering
  \includegraphics[width=16cm]{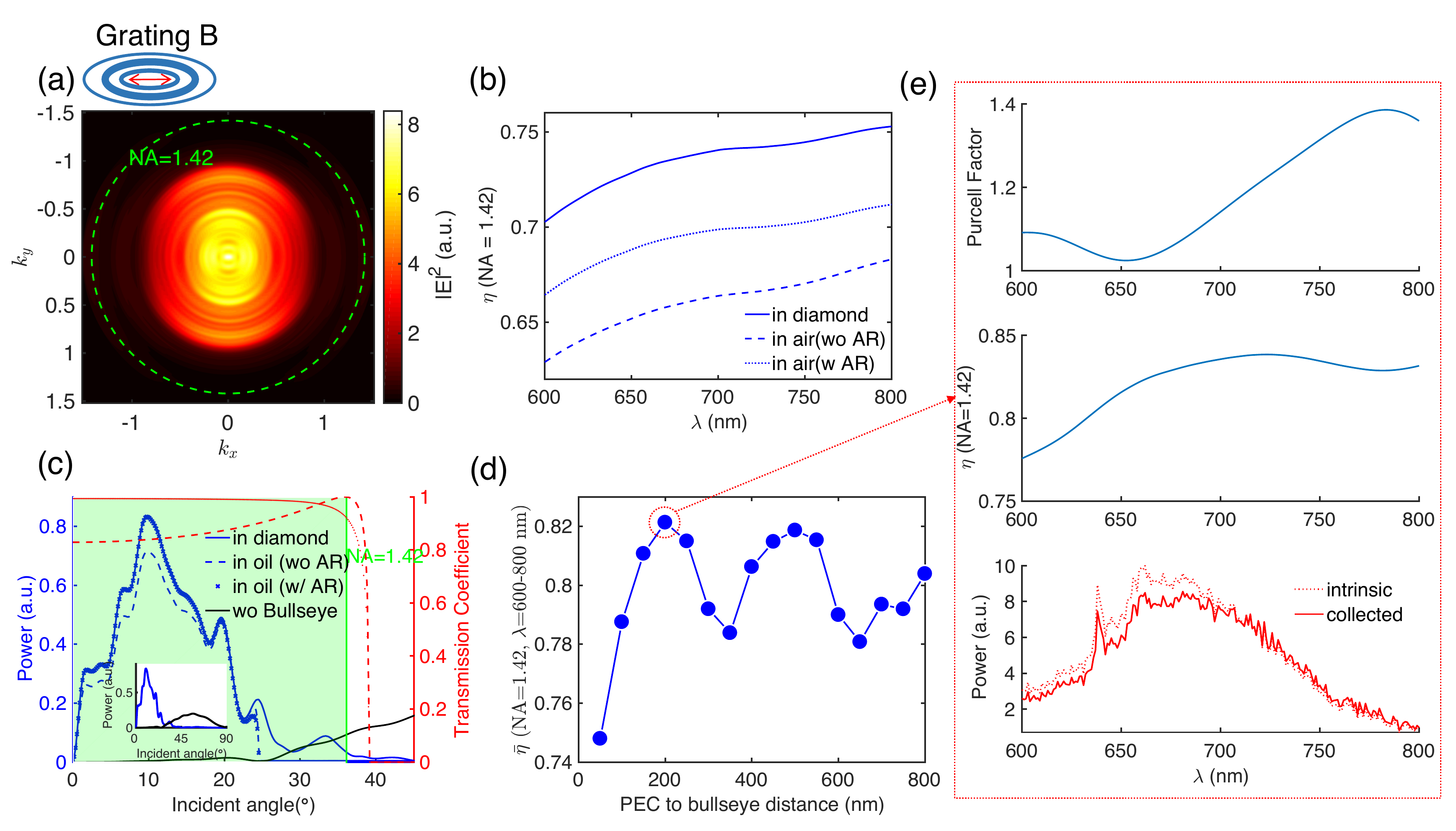}
\caption{\textbf{For Grating B:} (a). Spectrally integrated far-field pattern using oil collection. The dashed green circle indicates NA of 1.42. (b). Broadband collection efficiency $\eta$ with NA = 1.42 oil objective lens, comparing between the emission power in diamond (solid line), collected power in oil without (dashed line) and with (dotted line) an AR coating. (c). Left axis: Power (for $p$ polarization) angular dependance on the bottom diamond-air interface, comparing between the power in diamond (solid blue line), power collected in oil without (dashed blue line) and with (blue crosses) an AR coating. The green shaded area shows the angular extent of NA = 1.42. Inset: Power angular dependance over full 0-90$^\circ$ range. Right axis: Angular transmission coefficient for diamond-air interface without (red dashed line) and with (red solid line) an AR coating. (d). With PEC mirror over the grating at varying distance, maximal weighted collection efficiency $\bar{\eta}$=\bNVARPEC is obtained at 200 nm. (e). For 200 nm case in (d), broadband Purcell factor $F_p$ (upper panel), collection efficiency $\eta$ (middle panel) and convolved spectrum (${\eta} \times F_p \times P_{\text{NV}}$, solid line in lower panel), $P_{\text{NV}}$ is the intrinsic NV spectrum (dotted line).
}
\label{NVRT}
\end{figure}

The performance of the optimized grating design (Grating B) is shown in Fig.\ref{NVRT}. The far-field emission pattern shown in Fig.\ref{NVRT}(a) indicates that the broadband emission can be largely captured by a lens with an NA of 1.42. The unweighted collection efficiency within this NA is plotted as a function of wavelength in Fig.\ref{NVRT}(b), illustrating the broadband performance of this design. The wavelength-averaged weighted collection efficiency for this grating is $\bar \eta$ = \bNV, which is increased to $\bar \eta$ = \bNVAR when a thin-film AR coating optimized for 662 nm is added to the diamond-oil interface. For comparison, the average weighted power emitted within the same NA in diamond is $\bar \eta$ = 73.45\%. The benefit of the AR coating is further illustrated by comparing the angular distribution of the emitted power without [Fig.\ref{NVRT}(c), dashed line] and with [Fig.\ref{NVRT}(c), crosses] the AR coating. The collected power is enhanced over a large angular range ($\sim$ 0--24$^\circ$) and brought close to the intrinsic power measured in the diamond.

To further increase the collection efficiency by minimizing the power lost in the upwards direction, a PEC mirror is placed over the grating. As indicated in Fig.\ref{NVRT}(d), the optimal position of the mirror is 200 nm above the grating, which results in an averaged weighted collection efficiency of $\bar \eta$=\bNVARPEC, corresponding to a 6.5-fold enhancement relative to an NV in an unpatterned diamond film.

To verify that the optimized design is tailored for the input NV emission spectrum, we plot in Fig.\ref{NVRT}(e) the Purcell factor and collection efficiency $\eta$ over 600-800 nm range. Better performance is seen in the wavelength range where NVs emission is stronger. Note that the Purcell factor lies within the optimal range for maximal SNR.\cite{15PRBNVPurcellSNR}  The results show the advantage of the chirped circular grating for broadband operation, compared to conventional gratings\cite{15NanoLett.Li.MemBullseyeNV}. 

We also compare the intrinsic NV emission spectrum (indicated by the dotted red line, measured experimentally) with the predicted collected power (emission spectrum $\times$ collection efficiency $\times$ Purcell factor, solid red line), indicating that near-unity photon collection is possible with these gratings.

\subsection*{Tolerance of the chirped circular gratings to NV displacement}
\begin{figure}[htbp]
 \centering
  \includegraphics[width=8cm]{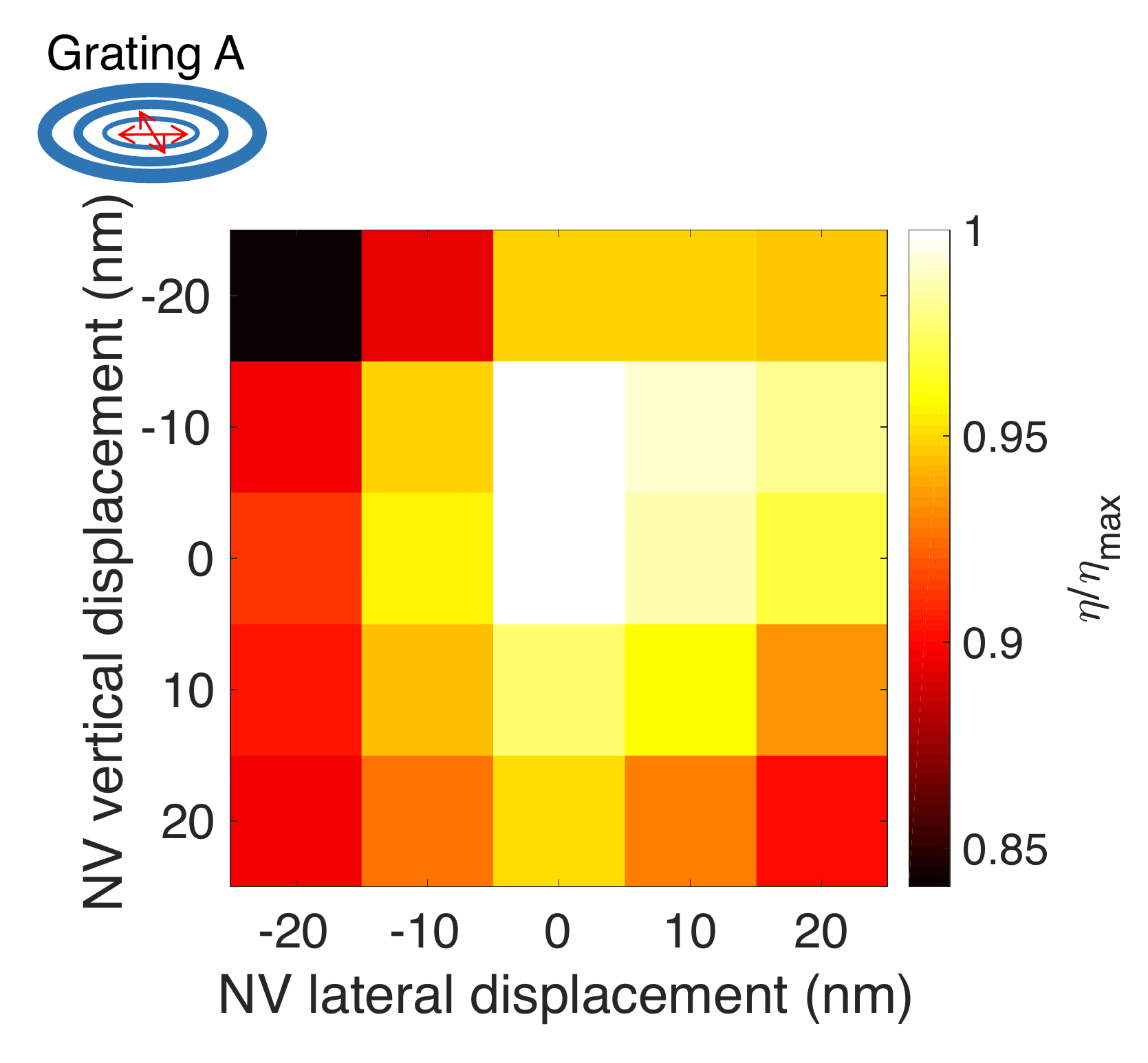}
\caption{ Effect of the NV's lateral and vertical displacement from the center on the collection efficiency. The values are normalized to the point with maximal collection efficiency ($\eta/\eta_{\text{max}}$)}
\label{FabTolerance}
\end{figure}

It is also important to consider the robustness of grating performance against the NV displacement from the grating center. 
The NV placement uncertainty for Nitrogen ion implantation is estimated from the statistical displacement range given by The Stopping and Range of Ions in Matter (SRIM) simulation~\cite{10SRIM}. For an NV depth of 100 nm, the lateral and longitudinal displacements are both $\sim $ 20 nm. We calculate the collection efficiency of Grating A (under the same conditions as scheme (a), without AR coating or PEC mirror) as a function of NV displacement. Here the emitter is modeled as NV with both $D_x$ along the [0$\bar{1}$1] direction and $D_y$ along [2$\bar{1}\bar{1}$] (a pair of dipoles whose contributions are summed incoherently, see Supporting Information S2 for the emission dipole orientations. This case is as also discussed by Sage \textit{et al.} \cite{12PRBNV4PortCollection}). The results in Fig.\ref{FabTolerance} show the $\eta/\eta_{\text{max}}$ versus the displacement in lateral and longitudinal directions, with $\eta_{\text{max}}$ being the collection efficiency for zero displacement. The results indicate that the collection efficiency for Grating A is within $\sim$ 15.9\% of $\eta_{\text{max}}$ under $\sim$ 20 nm of NV displacement. Such alignment accuracy can be achieved using the self-aligned ion implantation technique\cite{16APLpZhengNVnanobeamAlign, 17OMETimNVcavityAlign}.

%\section{Discussion}

\section*{Conclusions}
We proposed and numerically optimized chirped circular grating structures for near unity collection of NV fluorescence. The grating design is compatible with fabrication on a bulk diamond surface. Using an objective lens with an NA of 0.9, we show collection efficiency of \aZPLARPEC for NV ZPL emission, and averaged collection efficiency of \aPSBARPEC over the PSB range. For collection over 600 - 800 nm, we obtain weighted-averaged collection efficiency of \bNVARPEC into NA of 1.42 oil objective and moderate averaged Purcell factor of 1.18.  The detailed results are summarized in Table \ref{schemes}.
For the emitter-grating alignment, the collection efficiency varies 15.9\% for displacements within 20 nm from the center. Our approach is also applicable to emitters such as quantum dots or other solid state defects. 

The narrow emission angle of the chirped grating could also allow efficient coupling into high-NA single-mode fibers\cite{15PRappPrebleOnChipQI}. 

Furthermore, a dual grating-lens design where a second grating is added on the backside of the diamond film, can efficiently collimate the emission to certain target optical modes. Alternatively, a flat metasurface lens \cite{11.Sci.Yu.Antenna} could be positioned on the diamond backside for efficient interfacing to corresponding photonic structures.

Such grating structures can be fabricated using transferred hard mask lithography\cite{li2014nanofabrication}, or direct fabrication on the diamond using electron beam lithography or optical lithography\cite{16JOSABTimReview}.

\begin{suppinfo}
Discussion on the intrinsic NV$^-$ emission rate, dipolar component orientations of NV, estimation of baseline collection efficiency values for dipole near diamond surface, further simulation comparison with other structures, and limitations of grating depth are shown in Supporting Information. 
\end{suppinfo}

\begin{acknowledgement}

This research used resources of the Center for Functional Nanomaterials, which is a U.S. DOE Office of Science User Facility, at Brookhaven National Laboratory under Contract No. DE-SC0012704. The authors thank Tim Schr{\"o}der, Noel Wan and Matthew Trusheim for valuable discussions. J.Z. acknowledges partial support from the Office of Naval Research (N00014-13-1-0316). E.H.C. was supported by the NASA Office of the Chief Technologist's Space Technology Research Fellowship. D.E. acknowledges partial support from the NSF program EFRI ACQUIRE:``Scalable Quantum Communications with Error-Corrected Semiconductor Qubit'' and the AFOSR Optimal Measurements for Scalable Quantum Technologies MURI. 
\end{acknowledgement}



\bibliographystyle{unsrt}
\bibliography{BullseyePaperRef}

\end{document}

\newpage

\section{NV$^-$ intrinsic emission rate} \label{NVemissionrate}

\begin{figure}[htbp]
 \centering
  \includegraphics[width=12cm]{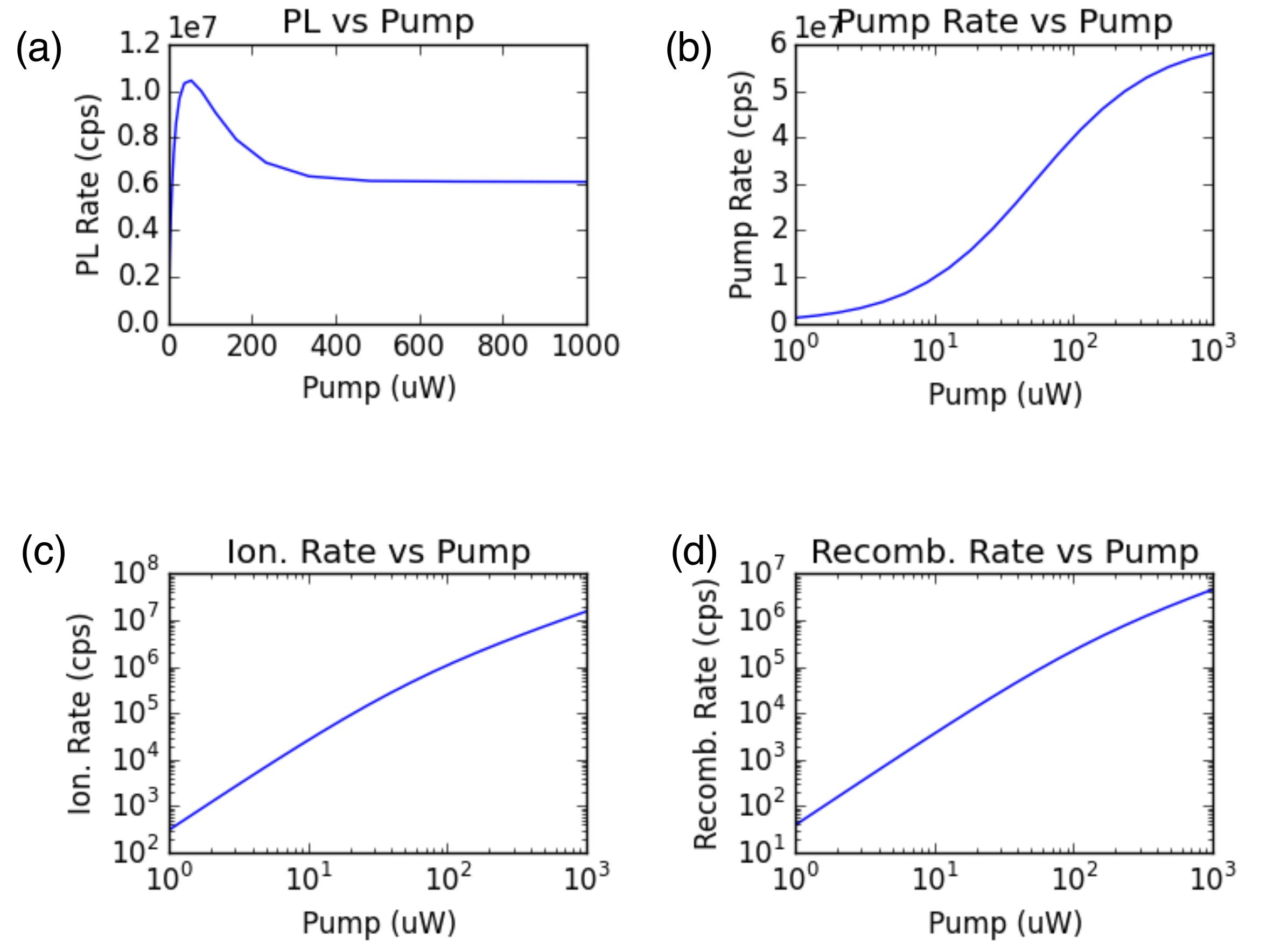}
\caption{(a) The simulated photon count rate from an NV as a function of laser pumping power; (b) The simulated NV pump rate as a function of laser pumping power; (c) The simulated ionization rate (rate of NV$^-$ transferred to NV$^0$ state) as a function of laser pumping power; (d) The simulated recombination rate (rate of NV$^0$ transferred to NV$^-$ state) as a function of laser pumping power.}
\label{NVUpperLimit}
\end{figure}

To accurately estimate the expected total NV$^{-}$ fluorescence rate, it is necessary to consider states other than the transition from the NV excited to ground state. Under continuous incoherent illumination, the electronic configuration of the NV$^-$ has a non-negligible probability of being ionized into a neutral charge state as well as spending a significant amount of time ($\sim$300 ns) in the long-lived metastable singled state. Both of these states contribute to a reduction in the ideal fluorescence rate of $\sim$ 1/13 ns $\sim$ 77MHz. Furthermore, the rates at which the NV transfers between these configurations change with the illumination intensity. Therefore, we use a four-level model to understand NV fluorescence under continuous illumination. Because most measurements of collection efficiency rely on the steady-state saturated fluorescence, we numerically calculated the steady-state populations of the four levels and estimated the continuous fluorescence from just an NV$^-$ as a function of illumination intensity. There are several important features to consider in the numerical results (See Fig. \ref{NVUpperLimit}). First, the ionization rate eventually becomes significant enough such that the total photoluminescence (PL) actually reduces (assuming no PL is collected in the NV$^0$ state). Second, if a saturation curve is improperly interpreted and extrapolated using the typical function ($\sim$ $R_{sat}/(1+I_{sat}/I)$), the total NV PL can be easily over-estimated. Finally, the numerically estimated saturated intensity from an NV can vary from 3 Mcps to 10 Mcps depending on the ambient condition of any particular NV leading to differing laser-induced ionization and recombination rates. These numerical results are consistent with previous experimental demonstrations \cite{12NJPEggelingNVdarkstate,15NLWrachtrupDiamondPost, 15NanoLett.Li.MemBullseyeNV}.

\newpage

\section{Collection efficiency estimation}  \label{CEestimate}
\begin{figure}[htbp]
 \centering
  \includegraphics[width=12cm]{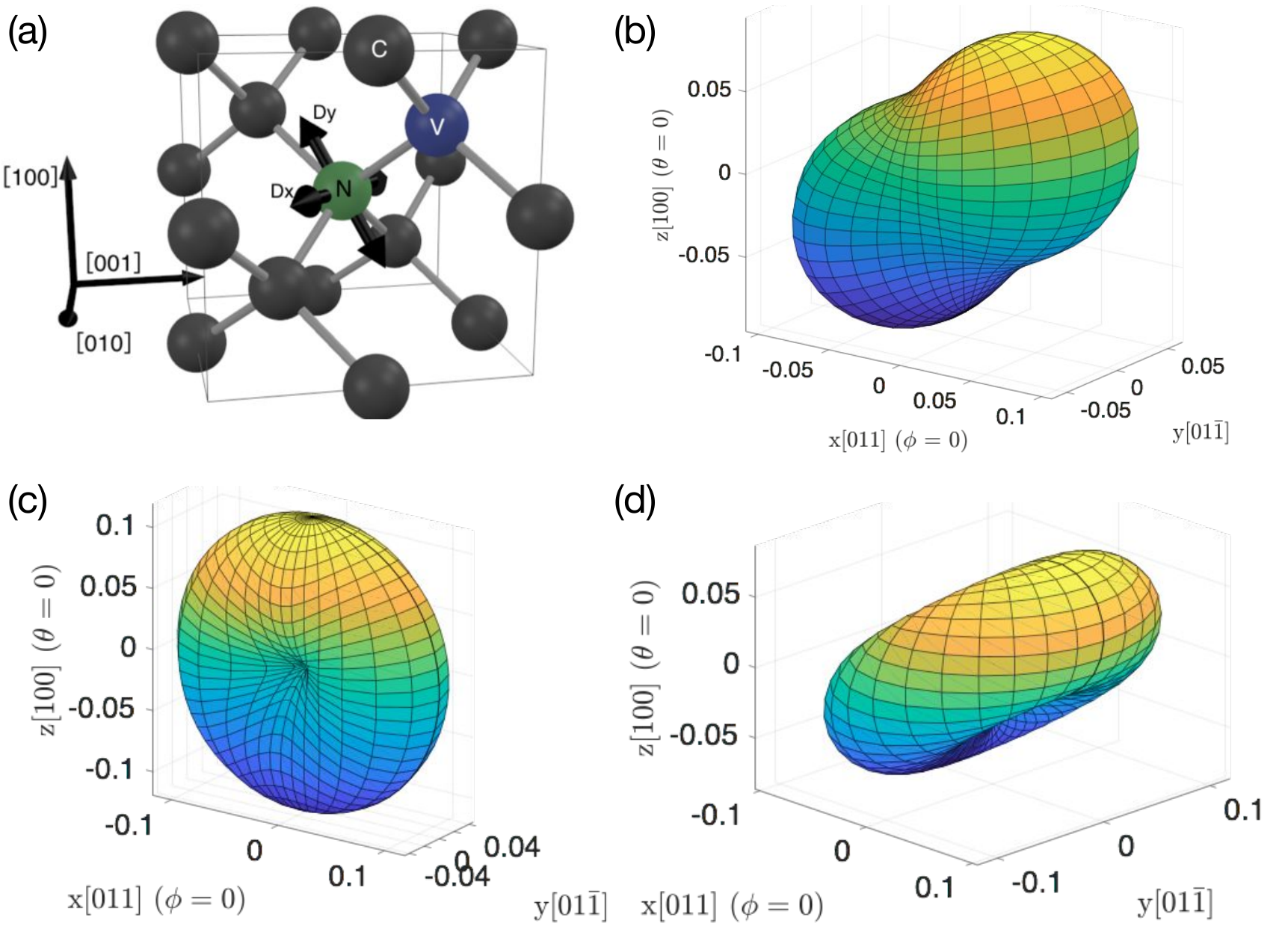}
  \caption{\textbf{NV dipole orientations and the 3D radiation profile}: (a). Atomic illustration of NV center structure, NV emission is from the two cross polarized dipole moment $D_{x}$ and $D_{x}$, both lying on the crystallographic (111) plane. (b) Radiation profile from the two dipoles with orientations indicated in (a). (c). Radiation profile for $D_x$ dipole. (d) Radiation profile for $D_y$ dipole. Viewing angle for the four subfigures are the same.}
  \label{NVStructure}
 \end{figure}

 As NV fluorescence is from two cross polarized emission dipoles, where their respective orientations depend on the direction of local strain field, collection efficiency can change for different terminating diamond crystal facets. Here shown in Fig.\ref{NVStructure}(a), the $D_{x}$ dipole is in the [0$\bar{1}$1] direction and $D_{y}$ dipole is in the [2$\bar{1}\bar{1}$] direction. Their radiation profile with respect to different diamond crystal facets are shown in Fig.\ref{NVStructure}(c) and Fig.\ref{NVStructure}(d).  For (100) terminated diamond, NV is pointing along [111] direction and has a polar angle of 54.7$^{\circ}$ (assuming the zenith in [100] direction), with two cross polarized dipoles lying in the (111) plane. The NV emission is an incoherent summation of $D_{x}$ and $D_{y}$ dipoles, as shown in Fig.\ref{NVStructure}(b). As the NV emission dipole orientations with respect to different diamond facets cannot be universally determined, we investigate the vertical and horizontal dipoles as two representative cases in this paper.
 
\newpage

Using dyadic Green's function \cite{novotny2012principles}, we examine the emission profile, collection efficiency into different numerical aperture (NA) and the Purcell factor for different dipole orientations at varying depth from the diamond surface. The Purcell factor calculation is also verified by FDTD simulations, showing good agreement between the two methods. The results show that collection efficiency can change more than 10 fold for different dipole orientations and more than 2 fold for different dipole depths.  This finding indicates that the collection efficiency needs to be carefully defined to compare the performance between different collection-efficiency-enhancing structures for dipole emitters.\cite{16LSA_Rishi_NVcollectionTaperFiber, 14JOAndersenNVSilverMirror, SILBrien, SILHadden, 11NJPBensonNDSIL, 14PRAppMaletinskySIL, 14BensonAPLParabolicMirrorND, 10NnanoLoncarNanoWire, opticalNVantenna, 15NLWrachtrupDiamondPost, choy2011enhanced, SiO2Grating, 15NanoLett.Li.MemBullseyeNV}

\begin{figure}[htbp]
 \centering
  \includegraphics[width=14cm]{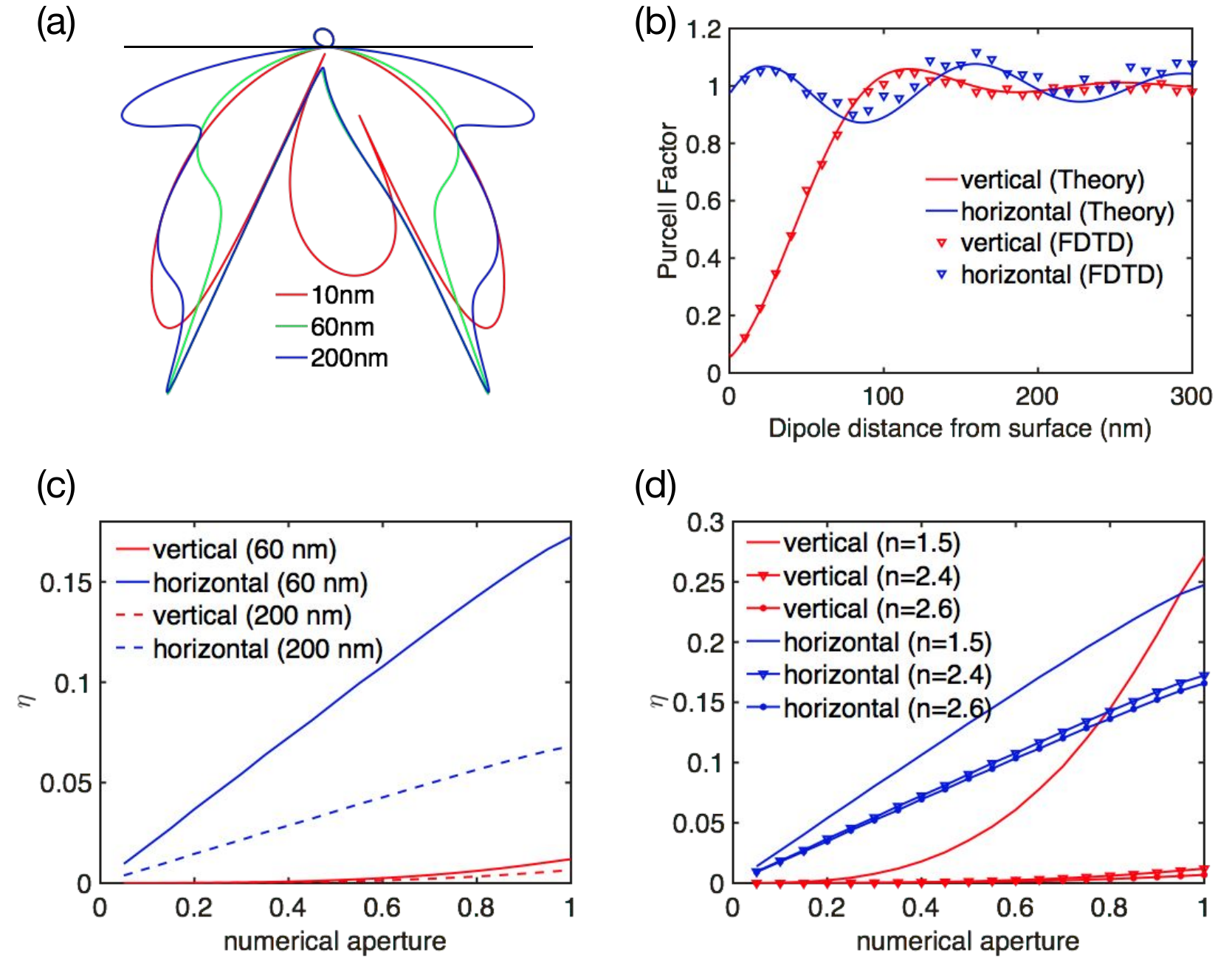}
\caption{\textbf{Theoretical calculations using dyadic Green's function:} (a). Far-field emission profiles of a dipole (assuming $D_y$) positioned 10 nm (red), 60 nm (green) and 200 nm (blue) below the diamond-air interface, which is indicated with the black solid line. (b). Purcell factor as a function of dipole depth from the diamond-air interface, solid lines are calculated using the dyadic Green's function, whereas the triangular data points are obtained from FDTD simulations; (c). Collection efficiency ($\eta_{\text{dipole}}$) as a function of numerical aperture. The dashed lines are for the 200 nm dipole depth, whereas the solid lines are for 60 nm deep dipoles. The red and blue color lines correspond to vertical and horizontal dipoles, respectively. (d). Collection efficiency as a fsunction of numerical aperture for different host material (n=1.5 (SiO$_2$), n=2.4 (diamond), n=2.6 (SiC)), The red and blue color lines corresponds to vertical and horizontal dipoles, which are 60 nm deep.}
\label{DipoleSurface}
\end{figure}

For dipoles in close vicinity to the diamond surface ($\sim$ $<$ 100 nm), it's spontaneous emission rate can change as a function of distance between dipole and surface. As shown in Fig.\ref{DipoleSurface}(a) where $D_{y}$ dipole is placed at different depth, the radiation profiles in air side stays the same while the emission patterns to diamond side are different. This indicates that the collected optical power in air side are identical, while the collection efficiency $\eta_{\text{dipole}}$, defined as $\text{P}_{\text{air}}^{\text{NA}}$/$(\text{P}_{\text{air}}+\text{P}_{\text{diamond}})$, will change as a function of dipole depth. 
To explore its dipole depth dependance, we calculate the Purcell factor for a dipole-near-interface model using dyadic Green's function. The results is displayed in Fig.\ref{DipoleSurface}(b), showing non-trivial variations for shallow dipoles ($<$ 100 nm), and negligible Purcell factor for deeper dipoles ($>$ 200 nm). The results are also verified using FDTD simulations, shown in Fig.\ref{DipoleSurface}(b) as scattered data points. 

The collection efficiency as a function of collection NA is plotted for horizontal and vertical dipoles in Fig.\ref{DipoleSurface}(c), showing the difference by dipole orientations and depths.
As evident from Fig.\ref{DipoleSurface}(d), refractive index of the host materials plays an important role to determine the baseline collection efficiency for their optical active defects. Data in Fig.\ref{DipoleSurface}(d) also provides an estimation for the collection efficiency from newly found defect centers in Silicon Dioxide\cite{16SciRepRabouwDefectSiO2} (n$\sim$1.5), Gallium Nitride\cite{17AMKwangYongGaN}(n$\sim$2.4) and Silicon Carbide\cite{16LienhardOpticaSiC}(n$\sim$2.6). In all cases, horizontal dipoles (plot in blue lines) show better collection efficiency into the air side, which indicates that (111) terminated diamond-air surface is favored for better collection efficiency where NV emission dipoles are horizontally oriented. To the best of our knowledge, there's no commercially available (111) terminated single crystal bulk diamond, and it is of great interest currently in the research community to obtain (111) terminated single crystal diamond using Chemical Vapor Deposition\cite{14APLWrachtrup111diamond, CVD111diamond}.

\begin{figure}[htbp]
 \centering
  \includegraphics[width=14cm]{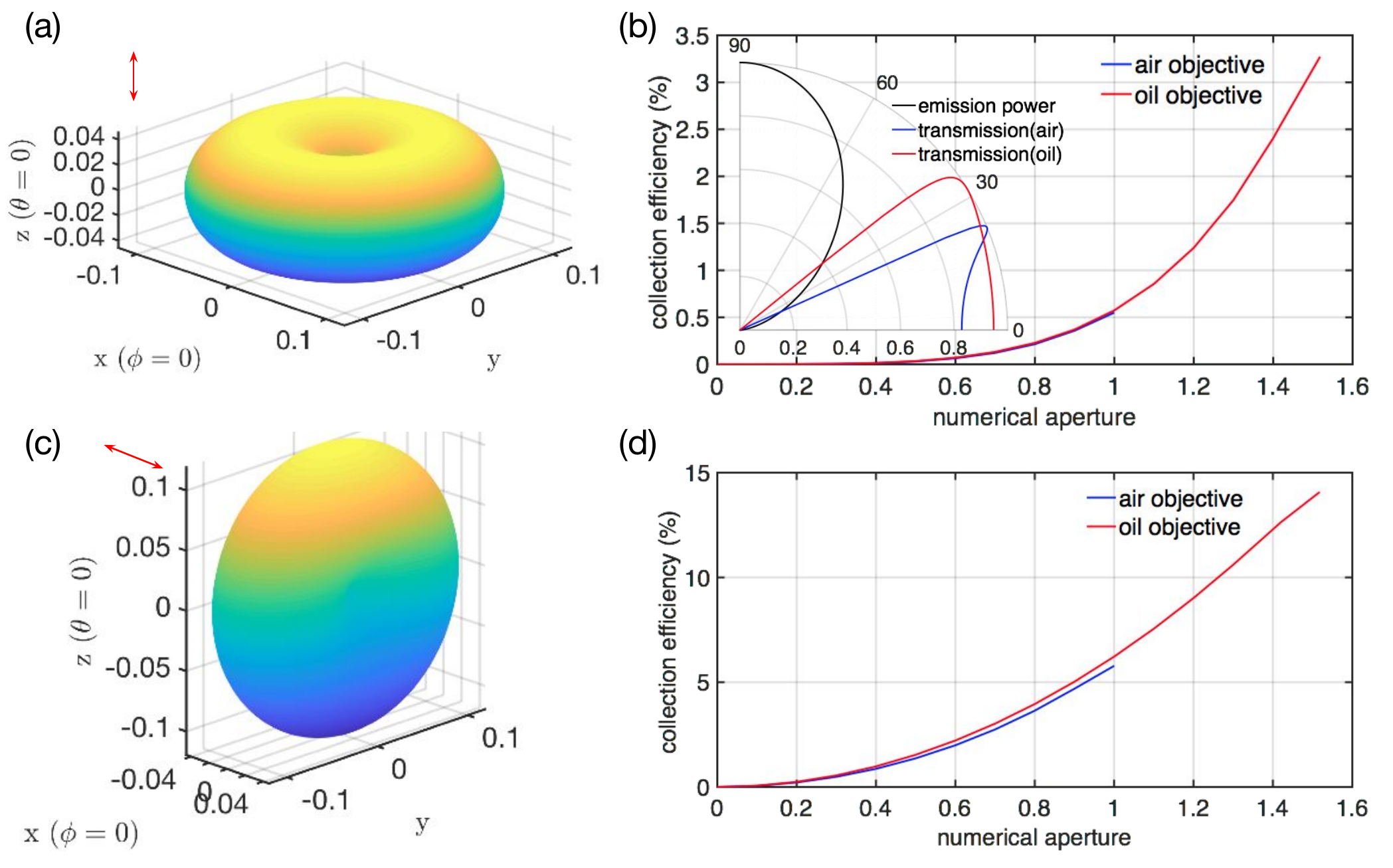}
  \caption{\textbf{Collection efficiency estimation based on wave optics:} (a). Radiation pattern of a vertical polarized dipole. (b) Collection efficiency as a function of numerical aperture, using air objective (blue) or oil objective (red). Inset: Polar plot of power versus polar angle $\theta$ (0-90$^{\circ}$), showing the intrinsic emission power (black, normalized to unity), collected power in air (blue) or oil (red).  (c) and (d). Same as (a) and (b), but for horizontal dipole. Note that no polar plot is shown in (d), since its emission profile is not circularly symmetric.}
  \label{DxDyNVemission}
 \end{figure}

For simplicity, we estimate the collection efficiency from wave optics and Fresnel's law for vertical and horizontal dipoles, assuming dry or oil-immersion objective collection. This provide the baseline value to compare with the results from the optimized circular gratings. The radiation profile for vertical and horizontal dipoles are shown in Fig.\ref{DxDyNVemission} (a) and (c). The corresponding estimated collection efficiencies for different NA are shown in Fig.\ref{DxDyNVemission} (b) and (d).  The results are quoted as $\eta_{\text{baseline}}$ in Table~1 of the manuscript. 

\newpage

\section{Results compared to previous reported diamond waveguide structures} \label{StuggurtPost}

From previously reported literature\cite{15NLWrachtrupDiamondPost}, a diamond waveguide structure is presented and experimentally verified to enable 1.7 Mcps with NA = 0.95 collection objective, NV is shallow implanted (5 nm deep). We take the optimal design presented (conically tapered nanopillar with top diameter $\sim$ 400 nm, bottom diameter $\sim$ 900 nm, and height of $\sim$ 1.2 um.)  The simulated results showed that the collection efficiency is $\sim$ 13-17\% and the Purcell factor is $\sim$ 1.  This results are in consistent with the previously calculated upper limit of photon count rate collected from a single NV$^-$ center in Fig.\ref{NVUpperLimit}, which is $\sim$ 10 Mcps.
\begin{figure}[htbp]
 \centering
  \includegraphics[width=12cm]{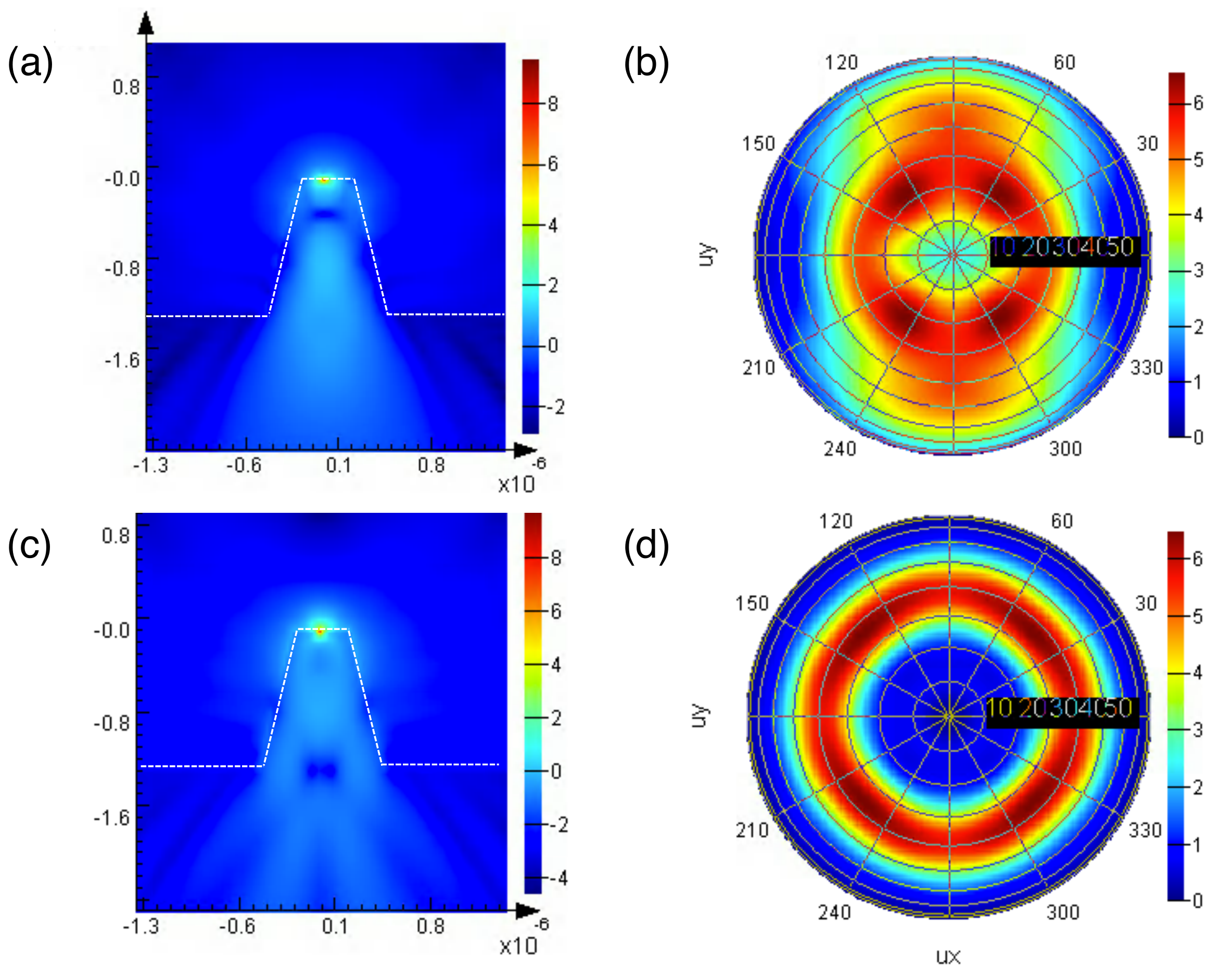}
  \caption{(a). $|\mathbf{E}|^2$ side view in $log_{10}$ scale, for in plane dipole. The dashed white line shows the outline of the diamond waveguide structure side view. (b). Far-field profile $|\mathbf{E}|^2$ for in plane dipole, for collection from the bottom side of the structure.  (c). $|\mathbf{E}|^2$ side view in $log_{10}$ scale, for out of plane dipole.  (d). Far-field profile $|\mathbf{E}|^2$ for out of plane dipole, for collection from the bottom side of the structure.}
 \end{figure}

\newpage

\section{Excitation limited Circular Grating etching depth}  \label{EtchDepthLimit}

\begin{figure}[htbp]
 \centering
  \includegraphics[width=10cm]{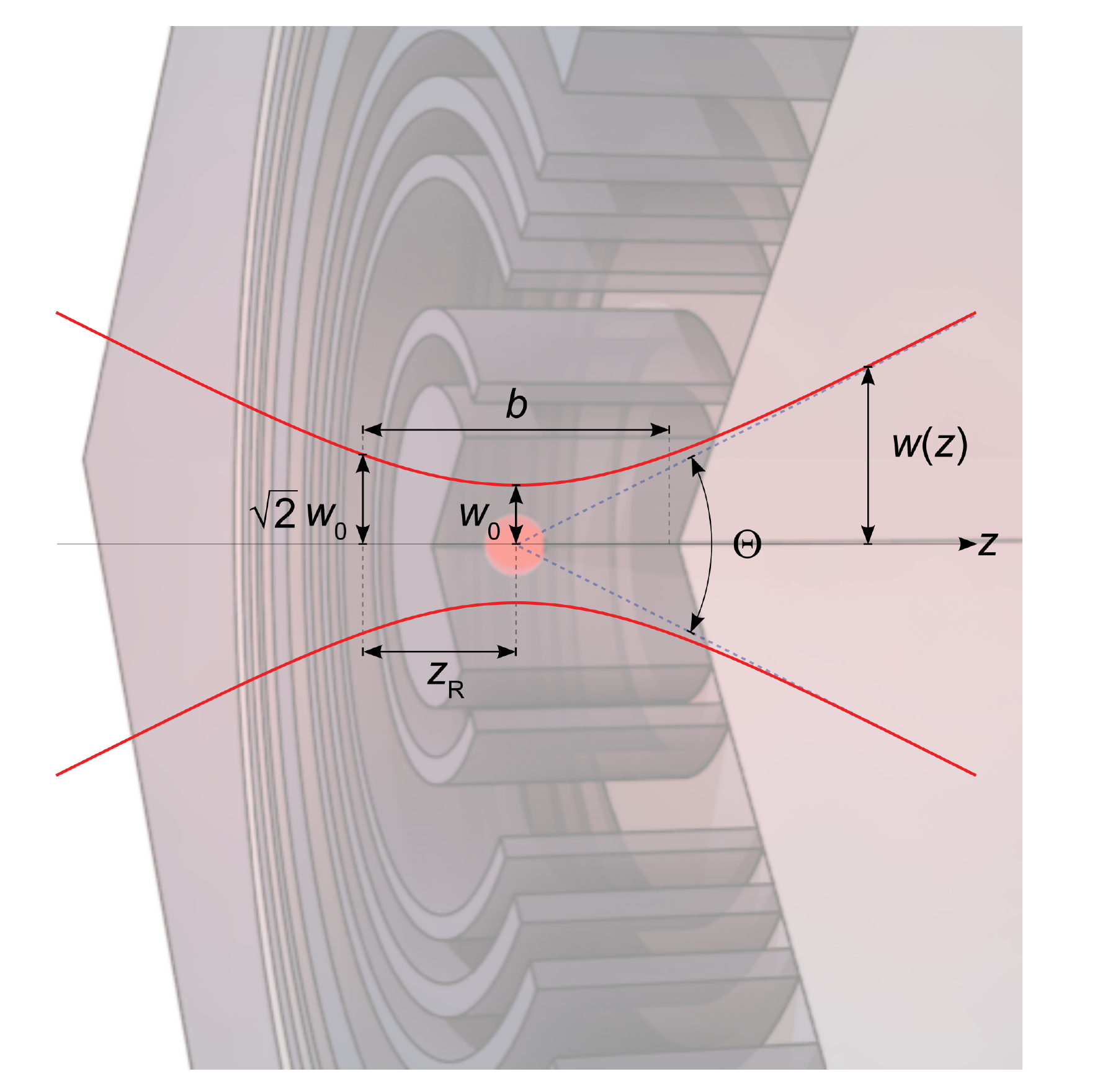}
  \caption{Illustration for optimal etching depth limited by the coherent laser excitation to minimize the scattering.}
  \label{ExcitationLimitDepth}
 \end{figure}

It's experimentally favorable to minimize the scattering between the laser excitation with the circular grating, especially when coherent pumping laser is used to drive the defect center \cite{16PanPRLPillarSPS}. This will lead to less background photoluminescence and significantly improve the measurement fidelity. Basically, the pumping laser represented by the Gaussian beam profile in Fig.\ref{ExcitationLimitDepth} will pose a limit to the height of the center post of the circular grating, to avoid the pumping laser to scatter with the lower edge of the center post. Such maximum etching depth is associated with the radius of the center disk and NV depth.

For a Gaussian beam with waist  $w_0$, the radius $r$ of the first diamond post is limited by $r > w(d_{\text{etch}}-z_{\text{NV}}) = w_0 \sqrt{1+(d_{\text{etch}}-z_{\text{NV}})^2/z_0^2}$, where $d_{\text{etch}}$ is the etching depth of the bullseye structure (also the height of the first post),  $z_{\text{NV}}$ is the NV depth and $z_0$ is the half focal depth of the Gaussian beam. For small beam divergence, $\text{NA} = n ~\text{sin}(\theta) = w_0/z_0$, therefore $r > r_{\text{min}} = \sqrt{w_0^2+(d_{etch}-z_{\text{NV}})^2\text{NA}^2}$ is the minimum radius for the first diamond post (with height $d$ and NV depth $z_{\text{NV}}$) in the bullseye structure, to minimize the scattering from coherent laser excitation through lens with numerical aperture of NA. In the limit of geometrical optics, where $w_0 \sim 0$, the minimum radius is simplified to $r > r_{\text{min}} = (d_{etch}-z_{\text{NV}})\text{NA}$, which agree with the intention to avoid scattering the excitation laser ray by the first diamond-air slit ring in the bullseye gratings.

\bibliographystyle{unsrt}
\bibliography{BullseyePaperRef}